# Frustrated Metastable Behavior of Magnetic and Transport Properties in Charge Ordered $La_{1-x}Ca_xMnO_{3+\delta}$ Manganites

Wiqar Hussain Shah[1,2], A. Mumtaz[3]

[1]Department of Physics, College of Science, King Faisal University,
Hofuf, 31982, Saudi Arabia

[2]Department of Physics, Federal Urdu University, Islamabad, Pakistan

[3]Department of Physics, Quaid-i-Azam University, Islamabad, Pakistan

## Abstract:

We have studied the effect of metastable, irreversibility induced by repeated thermal cycles on the electric transport and magnetization of polycrystalline samples of $La_{1-x}Ca_xMnO_3$ ($0.48 \leq x \leq 0.55$) close to charge ordering. With time and thermal cycling ($T$<300 K) there is an *irreversible* transformation of the low-temperature phase from a partially ferromagnetic and metallic to one that is less ferromagnetic and highly resistive for the composition close to charge ordering (x=050 and 0.52). Irrespective of the actual ground state of the compound, the effect of thermal cycling is towards an increase of the amount of the insulating phase. We have observed the magnetic relaxation in the metastable state and also the revival of the metastable state (in a relaxed sample) due to high temperature thermal treatment. We observed changes in the resistivity and magnetization as the revived metastable state is cycled. The time changes in the magnetization are logarithmic in general and activation energies are consistent with those expected for electron transfer between Mn ions. Changes induced by thermal cycling can be inhibited by applying magnetic field. These results suggest that oxygen non-stoichiometry results in mechanical strains in this two-phase system, leading to the development of frustrated metastable states which relax towards the more stable charge-ordered and antiferromagnetic microdomains. Our results also suggest that the growth and coexistence of phases gives rise to microstructural tracks and strain accommodation, producing the observed irreversibility.

PACS number (s): 75.30.Vn, 75.30.Kz, 75.60.Lr, 71.30.1h

Corresponding Author: wiqarhussain@yahoo.com

# I. Introduction:

The role of local charge-correlations and their competition with magnetic and electronic ground states is one of the most striking new features of the transition metal oxides such as high temperature cuprate superconductors and nickelates, but is most elegantly highlighted for the colossal magnetoresistive (CMR) manganites. The unusual properties of the CMR manganites involve, in general, a combination of charge, lattice, spin, and orbital degrees of freedom [1-6]. A typical formula for the perovskite-type CMR compounds is $A_{1-x}B_xMnO_3$ where $A$ represents a rare-earth atom (e.g., La) and $B$ an alkaline earth (e.g., Ca). For $0.2 \leq x \leq 0.5$ the materials undergo an insulator-to-metallic and paramagnetic-to-ferromagnetic transition. For $x>0.5$ the materials are, however, insulating and antiferromagnetic (AFM), and display a variety of charge and orbital ordering configurations. In the intermediate-doping regime (typically near $x=0.50$) the compounds can display both ferromagnetic (FM) and AFM and charge-ordered states, depending on the thermal and magnetic history and preparation conditions, e.g., oxygen or air ambient and preparation temperatures [7-12]. It has been observed that those prepared in oxygen show stronger AFM and insulating characteristics at low temperatures while the same compositions prepared in air show relatively more FM and metallic effects. The exact oxygen concentration affects the $Mn^{4+}/Mn^{3+}$ ratio as well as the prevalence of Jahn-Teller (JT) distortions. These latter have the effect of stabilizing the AFM and charge-ordered configurations. The subtle structural changes due to the JT induced distortions stabilize preferred orbital orderings and AFM alignments [14].

Phase separation in the manganites between hole-undoped AFM and hole-rich FM regions was theoretically predicted [13] and a variety of measurements [14, 15] have lent support to this scenario. For the $x=0.50$ composition both electron and x-ray diffraction measurements [3, 6, 16] have shown the coexistence of ferromagnetism and charge ordering in the form of microdomains. The FM-to-AFM transition is characterized by an evolution of the charge-ordered domains. The FM phase has itself been shown [7] to be spatially inhomogeneous consisting of both the incommensurate (IC) charge-ordered and FM charge disordered microdomains of size about 20-30 nm. The FM to AFM transition is characterized by an evolution of the IC charge ordered domains. Recent measurements [15] have also shown the presence of metastable states in the two-phase region of these compositions and the conversion of one phase into another when perturbed *externally*.

The main focus of attention in these materials is currently on the question of phase coexistence and separation, particularly close to the x=0.50 compositions [17, 18, 19]. The composition $La_{0.5}Ca_{0.5}MnO_3$ represents the boundary between competing (FM) and charge ordered anti-ferromagnetic (CO-AFM) ground states [17, 20]. At this boundary the FM-metallic state becomes unstable to an insulating CO-AFM state [20, 7]. The physical properties close to this phase boundary arise from a strong competition among FM double exchange interaction, and AFM super exchange interaction, and the spin phonon coupling [7-12]. There is considerable literature [20] on the transformation from a FM to an AFM (canted) and/or an insulator to a metallic state for compositions on the borderline of charge ordering. These transformations have been seen to be field and temperatures induced and are clearly *reversible*. After thermal cycling to above the charge ordering regime, subsequent cooling reproduces the initial states.

Our recent experiments reported here on charge-ordered compositions of $La_{1-x}Ca_xMnO_3$ ($x$=0.50 and $x$=0.52) manganites prepared in air and at slightly low temperatures ($T$<1400 °C) provide clear evidence for *irreversible* transformations within the low-temperature phase. Our magnetization, ac susceptibility and dc resistivity measurements show the as-prepared samples initially exhibiting a FM component at low temperatures, in addition to the AFM and insulating component expected to be dominant in this composition range. However, with time and/or thermal cycling the samples relax *irreversibly* into a highly resistive state with a pronounced decrease of the FM-metallic component. Our primary results show the presence of a two-phase system which has a metastable FM and metallic part. The charge-ordered compositions ($x$=0.48 and $x$=0.55) does not shows any irreversible and metastable behavior.
These results are remarkable in the sense that no such *irreversible* changes from one magnetic phase to another (or from metallic to insulating) have been observed in any of our compounds. All reported transformations with field or temperature have been reversible, and the initial state is recovered when the materials are re-cooled from high temperature. The primary effects we have observed are consistent with the formation of an initial metastable state with significant FM and metallic part. On cooling down below the charge ordering temperature there is apparently a growth of the AFM and insulating regions and a concomitant decrease of the FM and metallic ones, leading finally to a stable state after many hours of relaxation and/or repeated thermal cycling across the charge ordering temperature. These observations allow us insight into the dynamics of the conversion of one phase into the other.

The unique aspect of our work is the observation of these effects in a nominally "true" or unperturbed composition unlike many other cases where dopants lead to instabilities. We show that the metastable state at low temperatures gradually relaxes into the stable AFM and more resistive state with thermal cycling. Two main questions that arise in the context of phase separated systems are about the stability of the system once it has relaxed into a stable state in particular the conditions whereby the initial metastable state can be recovered. Secondly one would like to know the extent of the temperature range in which the (microstructural) changes accompanying the thermal cycling take place as the system converts from the metastable FM to the stable AFM phase. In particular, in the CMR systems under discussion, are the electronic changes confined only to the temperature region below the charge ordering temperature or extend above it? Our current work explores the metastable behavior in these directions. We explore the revival of the metastable state by high temperature annealing in a system that has relaxed to a stable state after undergoing a number of low temperature cycles and the evolution of this revived metastable state to a stable state. We also study the change in the relaxation dynamics as a function of (low) temperature cycling. These changes have been studied through DC and AC magnetizations, and magnetic relaxation at various temperatures.

## II. Experimental:

The samples reported here had the composition $La_{1-x}Ca_xMnO_{3+\delta}$ (x=48, 0.50 and 0.52, 55), were prepared in air by the standard solid-state reaction method, except for the slightly lower temperature of final anneal of 1275°C. Powders of $La_2O_3$, $MnO_2$, and $CaCO_3$ were finely ground and initially reacted at 1000 °C. The powders were reground and heated up to 1100 °C for 17 h in air. A final heat treatment was given to the pellets at 1275 °C for 17 h, again in air. The same procedure was adopted for the preparation of all compositions studied here. The x-ray patterns were studied for all the compositions. At room temperature all the compositions could be indexed to a tetragonal unit cell, as reported previously [21]. The in-plane lattice constants decrease from 5.460 to 5.409 Å as $x$ is varied from 0.35 to 0.52, while $c$ decreases from 7.730 to 7.662 Å over the same range. No unidentified peaks could be seen in any of the compositions, testifying to the structurally single phase nature of the materials, at least at room temperature as shown in the **Fig.1**. No structural changes were observed with the doping of different cations.

The oxygen stoichiometry of the samples was checked using the redox titration technique (iodometry). All the compositions were determined to be having excess oxygen. The excess oxygen parameterized by δ in $La_{1-x}Ca_xMnO_{3+\delta}$ was determined to be equal to +0.015 for both the x=0.50, 0.52 compositions. It is understood that excess oxygen can lead to the presence of vacancies on the cation sub-lattice placing the Mn-Oxygen and La (Ca)-O bonds under tension and compression respectively [18]. Furthermore, the presence of vacancies on the metal sub-lattice increases the concentration of $Mn^{4+}$ according to the relation $[Mn^{4+}] = 2\delta + x$ [22].

The zero field DC resistivity was measured by the standard four-probe technique in the temperature range 77<T<300K. The temperature dependence of Magnetization (M-T) was measured using the vibrating sample magnetometer. AC susceptibility measurements were performed on these samples and measurements were repeated several times, starting with an as prepared sample each time. The effect of thermal cycling on the AC susceptibility and the temperature dependence of the relaxation rate and the effect of high temperature annealing on the magnetic relaxation are measured.

## III. Result and Discussion:

To study thermal fluctuation mechanism we performed DC resistivity by the usual four-probe method, in the temperature range $77<T<300$ K. The main purpose of these measurements was to investigate the metal-insulator transition, conduction, and the most important is the metastable behavior of the compound with the repeated thermal cycles in compounds close to charge ordering compositions. We also needed to investigate the relation of the resistivity with the off-stoichiometric oxygen contents of the compounds. In all the following measurements, resistive as well as magnetic, it is important to distinguish between the response of the fresh samples and the samples that have been thermally cycled. In general the response in these states will be seen not to be the same. They will be referred to as *as-prepared* and *thermally cycled*, respectively.

To study the thermal cycling effects in $La_{0.50}Ca_{0.50}MnO_3$ composition, we have repeated the same sample several times from 77-300 K, as shown **Fig. 2**. The cure (a) is for the as-prepared sample and is taken while cooling down. Here the peak at *T*~141 K is evident, below which temperature the decline is consistent with weakly metallic behavior. On waiting at the lowest temperature 81 K, the resistance was observed to be increasing with time. The data labeled (b) in Fig. 1 show the subsequent resistive behavior on heating. The increased value of resistivity in (b) (at, e.g., the maximum) of 23 Ω-cm compared to the corresponding value for the initial state (a) of 18 Ω-cm is evident. The same sample was again cooled down from room temperature to 81 K and allowed to stay at the temperature for about 16 h. A very pronounced increase in the resistivity between 81 and 160 K is evident from curve (c) in the figure. The final value of the resistivity was ~ 38 Ω-cm. It is apparent that the low-temperature state becomes increasingly insulating and the peak in the resistance shifts to lower temperatures, with time and thermal cycling. It may be noted that these changes in the conducting behavior are *irreversible* in the sense that subsequent cool-downs from room temperature to low temperatures do not recover the initial low resistance state of **Fig. 2 (a).** Hence it is clear that the effect of thermal cycling has been to convert at least part of the metallic microdomains into insulating or highly resistive ones, leading to an overall increase of the resistivity. The $x$=0.52 sample (see inset of **Fig. 2**) displayed insulating behavior in the entire temperature range ($T$>77 K) even in the as-prepared state, and the resistivity was ρ ~ 4000 Ω -cm at 80 K. The changes in the resistivity with time and thermal cycling in this composition, though present, were much less pronounced (~ +10%) as compared to the $x$=0.50 composition.

To investigate the metastable behavior in other composition which is away from the phase boundary, we studied two other compositions with 48% and 55% Ca. Neither of these compositions showed any thermal cycling effects. The DC electrical resistivity measurements for $La_{0.52}Ca_{0.48}MnO_3$ were performed in the temperature range 77 to 300 K. The material shows a metal-to-insulator transition with a characteristic peak at the transition temperature 235 K. The sample is paramagnetic from ambient temperature down to $T_p$. Below the transition temperature $T_p$, the sample become metallic as well as ferromagnetic. The $La_{0.48}Ca_{052}MnO_3$ compositions were found to be insulating in the entire temperature range i.e. there is no peak in the resistivity and sharp rise in resistivity at 110 K.

**Magnetization:**

The DC magnetization studies were performed to identify how the *metastable* and *irreversible* changes observed in the resistivity were reflected in the magnetization, since the magnetic and resistive behavior are so intimately connected in these materials. The DC magnetization studies were performed under different conditions of the applied field and temperature i.e. field cooled (FC) and zero field cooled (ZFC) magnetization. For comparison purposes we also studied two other compositions with 48% and 55% Ca. Neither of these compositions showed any magnetic relaxation or thermal cycling effects. Thus it was clear from the absence of any metastable behavior despite the oxygen non-stoichiometry in these compositions, (x=0.48 and x=0.55) that both these composition have stable FM and AFM ground states respectively. This is understandable since both of these compositions are far enough from the phase boundary region where the FM and AFM/CO states are comparable in energy and hence are insensitive to the mechanical strains and the changes in the $Mn^{4+}/Mn^{3+}$ ratio produced by the excess oxygen.

The dc magnetization [$M(H)$] of these compositions was also studied. We have observed a magnetic moment of 2.44 $\mu_B$ /Mn and 0.97 $\mu_B$ /Mn for 50 and 52% Ca composition respectively with an applied dc field of 10 kOe, at 77 K. The same samples after repeated thermal cycling showed a pronounced decrease (25% and 18%, respectively) in the magnetic moments. The DC magnetization measurements as a function of temperature [$M(T)$] for the $x$=0.50 composition are shown in **Fig. 3**. The field-cooled (100 Oe) Measurements [Figs. 3(g) and 3(h)] showed a ferromagnetic transition at around 220 K and a slight decrease in the moment starting at about 125 K. Zero-field-cooled measurements [Figs. 3(k) and 3(l)], on the other hand, showed a very clear transition into the antiferromagnetic state at low temperature, with $T_N$ ~ 170 K. On thermally cycling the sample a significant (~10%) decrease in the value of the low-temperature moment, compared to the initial (as-prepared) value, was noticeable. A decrease of the moment was noticeable when waiting for long periods of time. However, in the presence of an applied dc field changes in the moment with time were too slow to be accurately determined by us. If the metastability comes only from the magnetic structure of the materials, i.e. spin canting, spin dynamics or due to spin frustrated structure, then with the application of high magnetic field, we expect to reduce or stop the process of metastability with repeated thermal cycling.

## Suppression of Metastable Behavior by Larger Magnetic Fields

The effects of thermal cycling on 50% Ca composition has been seen to enhance the metastable behavior and to convert at least part of the FM microdomains into AFM ones, leading to an overall decrease of the magnetic response. In general the AFM microdomains are understood to grow at the expense of the FM ones with thermal cycling. However the 50% Ca composition shows that at 77 K the material is neither fully FM nor fully AFM. There may be a canting of the spins or as is now more commonly understood there is a coexistence of both the FM and AFM phases at low temperatures. It is obvious that the inter-conversion of one phase into another depends to some extent on external factors such as number of cycles and temperature etc. These effects were explored further with the application of higher magnetic fields and later, as we shall discuss, with the high temperature anneals. Here we discuss the effects of magnetic fields on the metastable behavior. The rationale of the experiment is that since the material is transforming from a more FM to a lesser one, the effect of a field should be to slow down the conversion by somewhat stabilizing the FM subdomains. We applied a high DC field ($10^4$ Oe) to an as-prepared 50% Ca composition and recorded the temperature dependence of the magnetization in the FC mode as shown in **Fig 4**. As opposed to the behavior in low fields we observed no significant decrease in the DC magnetization even after several thermal cycles in this relatively high field. Subsequent to the thermal cycling in this high field (on the same sample) we decreased the field down to 100 Oe. The sample now displayed the instability, and the moment was observed to be decreasing with repeated thermal cycles, as observed for the low field cooled samples.

This illustrates the point we noted above, viz. the large DC applied magnetic field favors the FM phase and therefore prevents the system from converting the FM microdomains into the AFM ones. This means that the AFM/CO state is fragile while applying the field. This obviously implies that the effect of the high magnetic field is to stabilize the otherwise unstable or metastable ferromagnetic phase, thereby increasing the overall volume fraction of the ferromagnetic phase at low temperature, an effect described by F. Parisi *et al*. [23] as *ferromagnetic fraction enlargement*. F. Parisi *et al*. have shown that the effects of magnetic field in the two-phase system are dependent both on the magnitude and the mode of application of the field. In particular they discuss the ferromagnetic phase enlargement due to the application of the fields. We observed here that the fraction of FM and CO-AFM phases in 50% Ca composition can be controlled by appropriate thermal treatments, which produce a continuous change of the relative fraction of the competing FM and CO-AFM phases.

## DC Magnetic Relaxation for $La_{0.50}Ca_{0.50}MnO_{3+\delta}$

By cooling these metastable systems in a field or applying a field subsequent to cooling, we expect to see different degrees of conversion and metastable behavior. We studied the time dependence of the conversion from the FM to AFM in the subdomains. Towards this end we applied a DC field of 50 Oe from an electromagnet while cooling an as-prepared sample. Reaching the lowest temperature (80 K) the DC field was shutdown and we recorded the remnant magnetization data at this temperature for one hour at 5 seconds intervals. The magnetization decreases with time as expected. The data was logged onto a computer interfaced with the magnetometer. This is shown in **Fig. 5.** As is customary for relaxing systems with a distribution of relaxation times or barriers we fitted the M(t) data to the logarithmic form,

$$M(t) = M(0)[1 - r\ln(1 + \frac{t}{\tau})] \qquad 1$$

Here *M(0)* is the magnetization at time $t=0$, $\tau$ is a fit parameter called the *reference time* [**25**]. The magnetic viscosity or relaxation rate *r* was determined from the fit. Typical values obtained for this composition and temperature were in the range $r = 0.088 \pm 0.004$ as identical with the literature [**26**]. It is to be noted that this decrease of the moment is attributed by us to the same phenomenon viz. conversion of the FM to the AFM part and not to the more common relaxation process in any system with competing interactions or randomness of exchange. While the former are *irreversible* the latter are reversible as e.g. in a spin glass where cycling the temperature to above $T_g$ retrieves the initial state. In our case, thermal cycling after the relaxation did not recover the initial state.

The relaxation of the moment for the ZFC case was also studied. Here the fresh sample was cooled down to 80 K in H=0 and then a field (50 Oe) were applied. The field was held on for 5 minutes and was subsequently turned off. The moment was again seen to decrease with time albeit much slower than for the FC case. This is explainable as when cooled in zero fields the system is essentially having a larger AFM/CO fraction and the smaller FM fraction has a small amount of relaxation (towards the AFM/CO state). This small amount of relaxation is further suppressed by the application of the high DC field. On the other hand in the case of FC the fraction of the meta-stable FM state is initially large, i.e. the moments are trapped in the metastable FM state to a greater extent, and subsequently there is a pronounced relaxation.

## $La_{0.48}Ca_{0.52}MnO_{3+\delta}$ (LCMO-52)

We have studied $La_{0.48}Ca_{0.52}MnO_{3+\delta}$ sample in detail for the DC magnetization as a function of temperature in FC and ZFC modes with 100 Oe probing field. We have observed the decrease in the magnetic moment of the sample with thermal cycling (i.e. cooling the sample to 98 K and then heating to room temperatures and repeat this several times) both in the FC and ZFC mode. The data of **Fig. 6** were taken during heating the sample, prior to which the sample was cooled in zero applied magnetic fields. Curve (a) was taken when the sample was fresh (as prepared), (b) was taken after two thermal cycles and then curves (c) and (d) consecutively one after other. The instability in the low temperature magnetic state was observed. **Fig 7** is the FC data for the same sample where the two curve (a) and (b) are separated by several thermal cycles. Thermal Hysteresis effects (differences between cooling and heating data) were very pronounced for this sample. The data shows that the magnetization of the sample increases steeply with the decrease in temperature both in the FC and in ZFC modes and peaks at ~212 K. In ZFC there occurs a 20% decrease in the moment when the temperature is decreased down to 170 K but below this temperature the moment *increases* with further decrease in the temperature. In contrast for the FC case (Fig. 4.12) the moment instead of decreasing becomes constant in the temperature range of $170<T<212$ K while (as in the case of ZFC), it subsequently increases with further lowering of the temperature. At the same time, the effect of the thermal cycling is to decrease the magnitude of the magnetic moment at low temperatures, both in FC and ZFC data. The extent of the thermal cycling induced decrease in magnetization is however significantly less for the FC than for the ZFC case. For the case of ZFC all the curves corresponding to different thermal cycles coincide for temperatures above 200 K while for the case of FC the data coincide at 220 K.

## $La_{0.45}Ca_{0.55}MnO_{3+\delta}$ (LCMO-55)

The $La_{0.45}Ca_{0.55}MnO_{3+\delta}$ composition did not show any magnetic relaxation or thermal cycling effects. However differences between ZFC and FC were evident in this composition. Thus it was clear from the absence of any metastable behavior despite the oxygen non-stoichiometry in this composition, (x=0.55) that this composition has stable AFM ground state. This is understandable since this composition is far enough from the phase boundary region where the FM and AFM/CO states are comparable in energy and hence are insensitive to the mechanical strains.

The typical data of DC magnetization as a function of temperature for a fixed DC (100 Oe) field, of the sample LCMO-55 is shown in **Fig. 8**. In this figure three curves (a), (b) and (c) have been shown. The data labeled as (a) was taken when the sample was cooling in field where as the data shown in (b) was obtained when the sample was heating from 98 K to room temperature after (a). The data of plot (c) was recorded in the ZFC condition. There occur sharp increase in the FC magnetization for $T<270$ K and continues down to 225 K. The increase in M is very slow between $188<T<225$ K. The maximum magnetization in case of FC data occur at T=188 K and after this temperature the magnetic moment decreases all the way down to lower temperatures. This shows that this composition has dominant AFM interactions at lower temperatures (below 188 K). While heating the sample the maximum of the magnetization shifts to a higher temperature (218 K) while the maximum value itself decreases. The data shows a pronounced thermal hysteresis effect. Thermal hysteresis in this case is very large i.e. 40% as compared to other compositions, studied here. In FC, the maximum magnetization in heating is 29% less than the corresponding value for the data taken in cooling the sample at 230 K. The ZFC magnetization is overall smaller than the FC one, which again indicates that the AFM interactions are dominant in the low temperature state. While both the FC (heating) and ZFC data show a peak in M(T) at the same temperature (218 K), the peak in the case of ZFC is much sharper than from FC. There is no thermal relaxation observed with repeated thermal cycling from 77-300 K. This clearly indicates that the composition has moved away from the region where the two phases FM and AFM have very nearly the same energies at low temperatures and one of the phases (AFM) is more favored. As the calcium content is increased, the behavior goes from a FM at 48% to AFM at 55% passing through a region of mixed phase behavior for 50 and 52 % Ca. As expected the 50% is predominantly FM with some AFM clusters while the 52% is predominantly AFM and insulating with some FM clusters at low temperatures.

**Dynamic relaxation by AC susceptibility measurements:**

The phase dynamic or metastable behavior has also been studied using the AC technique. This had the advantage that the response in small ac fields could be studied. Furthermore the AC measurements being dynamic in nature are sensitive to the loss mechanisms and relaxation processes. The latter show up in the out of phase parts of the susceptibility. AC susceptibility studies of $La_{1-x}Ca_xMnO_{3+\delta}$ family of compounds for $0.48\leq x \leq 0.55$ were conducted using a self-made AC susceptometer.

We measured the magnetic relaxation for an applied AC field of 5 Oe in an as-prepared x=0.50 sample which was cooled rapidly down to 80 K by inserting it directly into liquid nitrogen. It was observed that the relaxation continued unabated for as long as nine hours. However, the rate of relaxation does decrease for long times, $t\geq2$ hours. The data shown in **Fig.9** are for a period of

1 hour. The AC susceptibility was observed to decrease systematically with time and was fit to a log function,

$$\chi'(t)=\chi(0)[1-r\ln(1+\frac{t}{\tau}] \qquad \ldots\ldots\ldots\ldots\ldots 2$$

Here $r$, $\chi(0)$ and $\tau$ are parameters obtained from the fit. It is evident that the fit is quite reasonable. From the fit the rate of relaxation at 80 K a value of $r$=0.087 (**left panel of Fig.9**) was determined. Another as-prepared sample was taken and was inserted into the cryostat, pre-cooled to 90 K. The logarithmic fit of this data is shown in the right panel of **Fig.9**. The relaxation rate obtained from this fit was $r$=0.047. We also measured the relaxation at 125 K on another as-prepared sample. The ensuing relaxation was seen to be considerably noisier and most probably reflects the fact that the two phases are now closer in energy unlike the low temperature region where the AFM is much stable. Hence both forward and backward jumps may well be occurring between the two potential wells corresponding to AFM and FM aligned spins. The value of the relaxation rate at this temperature as obtained from the fit was seen to be $r$= 0.032. The rate of relaxation clearly decreases with increasing temperature since that brings the sample deeper into the stable region of the FM state. The rate $r$ was also seen to decrease with time and thermal cycling. The latter observation indicates that as the sample relaxes towards the more stable low temperature state (AFM/ CO) with thermal cycling, the rate of conversion of the remaining FM regions becomes slower. Logarithmic relaxation has previously been observed [8, 20, 21] and is generally attributed [22] to a distribution of energy barriers separating local minima that correspond to different equilibrium states.

**Retrieved of the metastable state Annealing:**

We showed that the metastable state at low temperatures gradually relaxes into the stable AFM and more resistive state with thermal cycling. Two main questions are posed by the metastable behavior observed in these phase-separated systems. Firstly if such a system has relaxed into its stable state can it be made to revert back to a metastable state by appropriate thermal treatment or by some other means? Secondly one would like to know the temperature range up to which the (microstructural) changes take place as the system converts from the metastable FM to the stable AFM phase. We explore the *revival* of the metastable state by high temperature (800 $^{o}$C) annealing in a system that has relaxed to a stable state after undergoing a number of low temperature cycles and then follow the evolution of this *revived* metastable state to a stable state with further thermal cycling. We also study the change in the relaxation dynamics of the revived state as a function of (low) temperature cycling. These changes have been studied through DC and AC magnetizations, measurements and magnetic relaxation at various temperatures. We present data that provides more extensive information about the irreversible transformations within both the low (CO) and high temperature (FM) region; the end of the transformations with cycling; their revival with annealing and finally, the subsequent temporal behavior.

The conversion of the ferromagnetic-metallic state to one which is more resistive and anti-ferromagnetic was seen to be irreversible i.e. once the sample fully relaxes, the meta-stable state cannot be achieved by giving thermal cycle to the material from room temperature to 77 K and vice versa. However this meta-stable state could be recovered in samples that were fully relaxed (as a consequence of a large number of thermal cycles) by subjecting them to a high temperature annealing (700-900°C) for 8 hours. We annealed our samples both in air and in oxygen ambient and the unstable low temperature state (FM-metallic) was again recovered, irrespective of the ambient. These features are illustrated by the data of **Fig. 10**. The inset of Fig. 2 shows the AC susceptibility of the x=0.50 sample measured in successive heating cycles, prior to the annealing. On thermal cycling the change to a less ferromagnetic state below T<180 K is apparent. The peak at T~125 K is also noticeable. Here we focus on the change in the susceptibility as the sample is annealed to high temperature as described above. The resulting response is shown by the two sets of data (f) and (g) in the main figure of Fig. 2. This response is to be compared to the last set of data in the pre-annealed state (curve (e) in the inset). Between curves (e) and (f) the moment clearly increases very significantly (by about 50%) and also indicates an upturn below 110 K. Both these observations support the conclusion that annealing has reactivated or recovered the unstable ferromagnetic state. With low temperature thermal cycling of the annealed sample, the moment further relaxes towards the less ferromagnetic state (see curve (g) in Fig. 2) as in the un-annealed sample. Careful measurements on the annealed sample also made clear that successive low temperature thermal cycles had subtle effects in the region of the shoulder at 200 K. With cycling, the moment at the 200 K shoulder became slightly reduced.

## IV. Conclusions:

We concludes that the changes in the resistive and magnetic state of our system shows a metastable phase at temperatures below where the charge ordering and AFM transitions would be expected to take place. This is in line with the phase separation scenario where inclusion of the long range coulomb interaction promotes charge ordering. In the x=0.52 composition it is clear that even in the low temperature state observed (T<160 K), where FM correlations are initially dominant, there is no tendency for lowering of the resistance, and the insulating state persists. Thus in this state there are both FM and AFM regions, but the conduction process is still dominated by the AFM charge ordered clusters. In the x=0.50 composition on the other hand the FM correlations are accompanied by a metallic behavior below 130 K. With increasing time the changes in the resistivity and the decrease in the FM cluster are consistent with the transfer of electrons from some of the $Mn^{+3}$ ions to some of the $Mn^{+4}$ ions such as to make more of the system charge ordered and insulating. In other word more of the FM microdomains convert into AFM and charge ordered regions. It is noticeable that the changes in the susceptibility only occur for the region below T∽200 K, where the Jahn-Teller effects create the lattice distortions which are understood to stabilize the charge ordered state. The magnetic relaxation experiments give further confirmation of the relation between the passage through the temperature region close to $T_{CO}$ on the one hand, and the initiation of the transformations or phase dynamics, on the other. It would appear that the metastable FM state we observe is created due to less than optimum oxygen stoichiometry $La_{0.5}Ca_{0.5}MnO_3$ that is an outcome of annealing in air. The less than optimum oxygen value leads to an average concentration of $Mn^{+4}$ slightly less than x=0.50 and thus to a partially FM and (for x=0.50, a metallic) low temperature state. This does not allow the system to stabilize the charge ordered microdomains initially. It is thus clear that oxygen non-stoichiometry in plays a crucial role in determining the low temperature state of the border-line charge ordered compositions. In a generalized sense the existence of metastable states in such systems and glassy relaxation effects are consequences of the competing interactions and configurations, and a landscape of closely spaced energy states separating the coexistent phases.
The recovery of the initial meta-stable state by annealing the material at temperatures well below the reaction temperature is one of our important observations. To understand this we recall that the low energy state for these compositions (below T~160-180 K) is the CE-type magnetic structure, which is pinned into a meta-stable ferromagnetic state probably due to the mechanical strains in the lattice [14]. The strains arise from the mismatch of the lattice of the two phases. The non-Stoichiometric (excess) oxygen may also affect the strains through vacancy formation [11, 12]. We interpret the reactivation of the FM phase on high temperature annealing as suggesting that the mechanical strains, which may have relaxed by the repeated (low temperature) thermal cycling, are re-introduced into the lattice. Furthermore new oxygen vacancies may have also been introduced in the system with subsequent changes in the $Mn^{4+}/Mn^{3+}$ ratio. The net result is that phase separation and metastable tendency are further strengthened in the system. The above measurements indicate that the electronic and magnetic short-range correlations are undergoing transformations with thermal cycling below $T_{CO}$. We suggest that the thermal cycling not only has a pronounced effect on the long range order below $T_c$ and $T_{CO}$ but also has a significant effect on the short range correlations. In the samples that are re-activated ferromagnetically may be indicative of the formation of local ferromagnetic correlations.

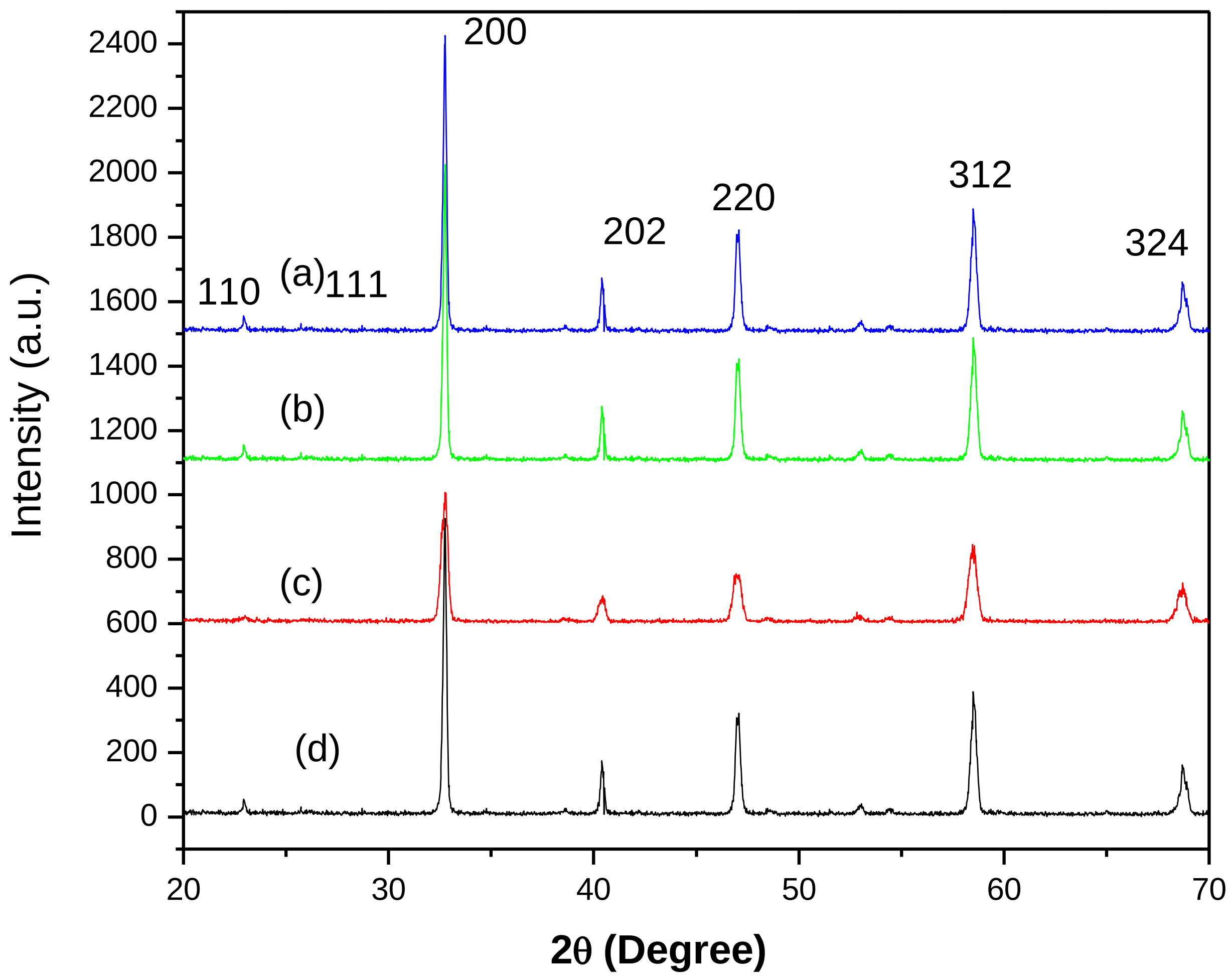

**Fig. 1**. X-ray diffraction patterns for $La_{1-x}Ca_xMnO_3$ are shown, where data (a), (b), (c) and (d) represents the x=0.48, 0.50, 0.52, 0.55 Ca concentration respectively.

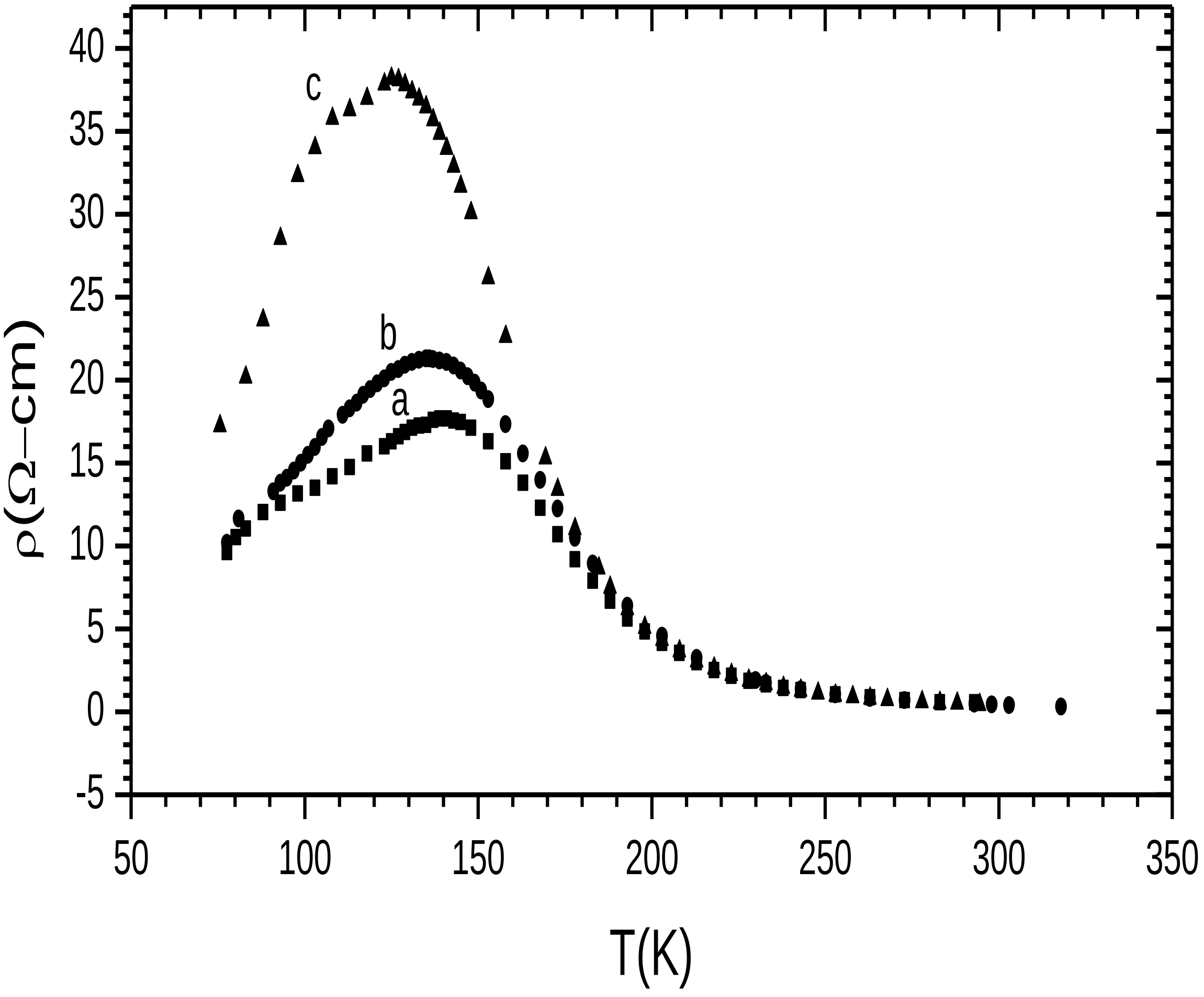

**Fig. 2.** Main Figure: Temperature dependence of the resistivity for $La_{0.50}Ca_{0.50}MnO_{3+\delta}$ Data labeled as (a) was taken during cooling of a fresh sample; (b) obtained on heating after completion of (a); while (c) were obtained subsequent to the (a) and (b) measurements during heating, after waiting at 80 K for 16 hours. The pronounced increase at low Temperatures is evident and was irreversible. Inset: Resistivity of x=0.52 sample as a function of temperature.

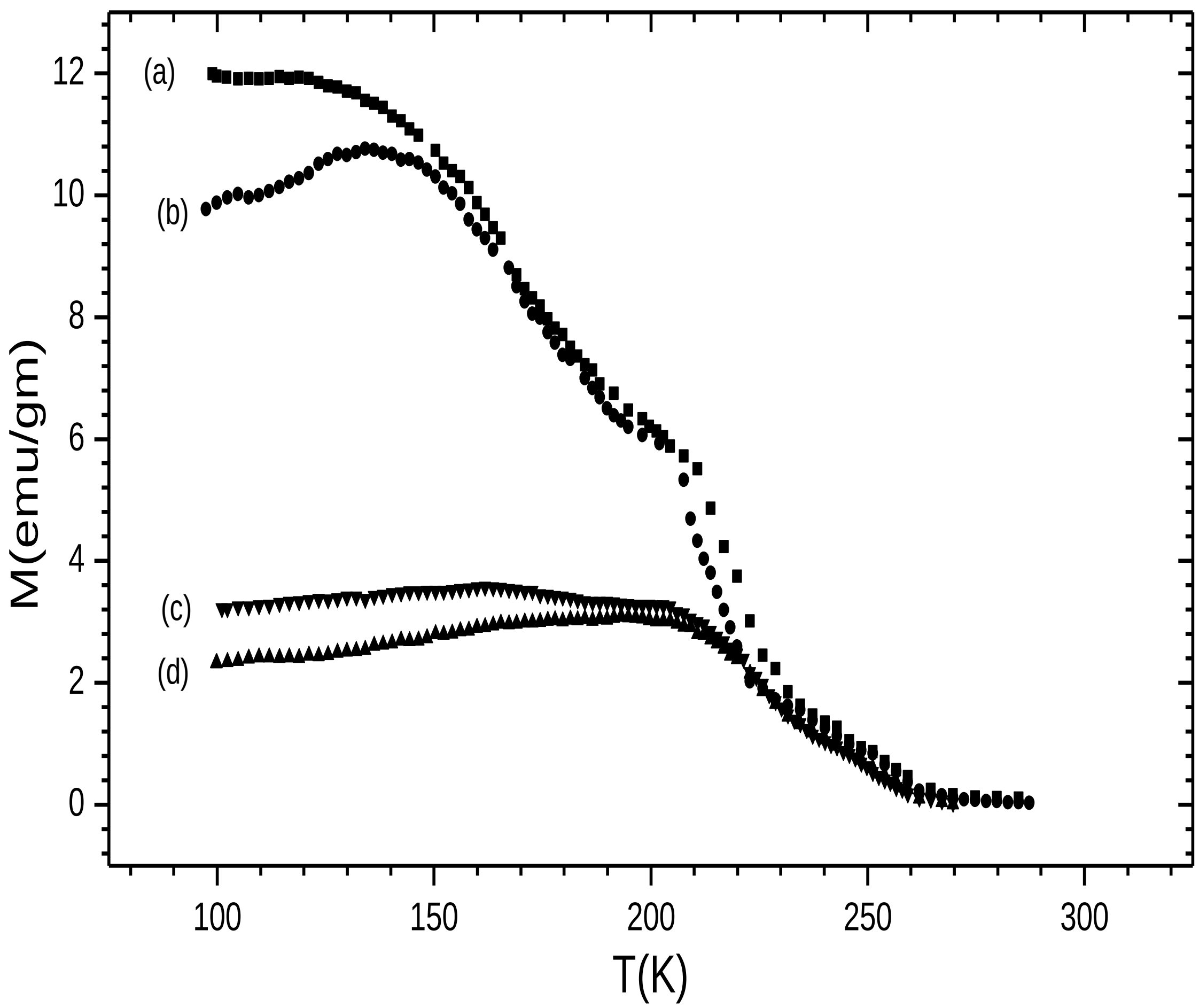

**Fig. 3.** Temperature dependence of DC magnetization of $La_{0.50}Ca_{0.50}MnO_{3+\delta}$ sample; ((g ) and (h) are field cooled while (k) and (l) are zero field cooled). Irreversible decrease in the moment from (a) to (d) is evident.

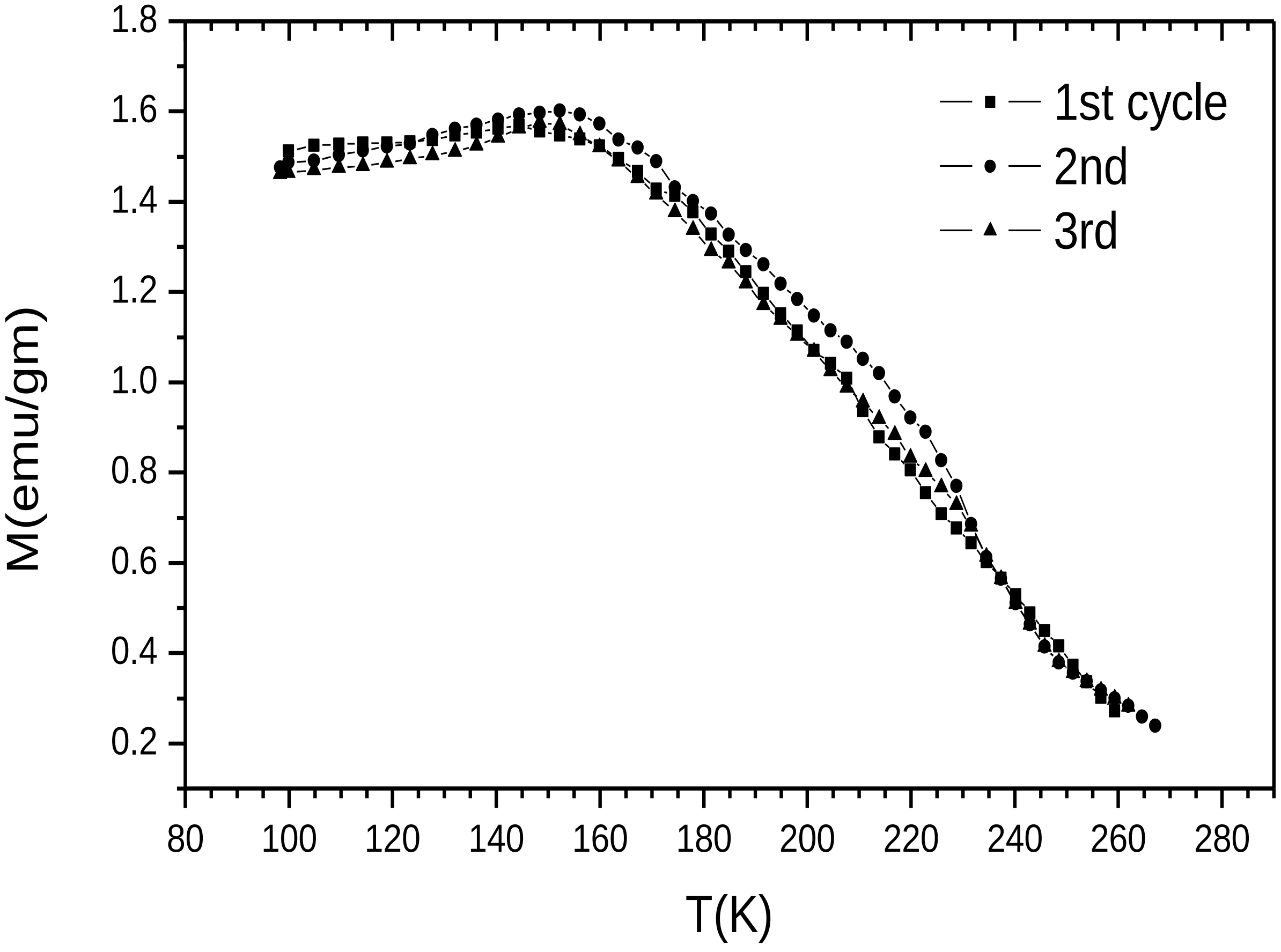

**Fig. 4**. Temperature dependence of DC magnetization for $La_{0.50}Ca_{0.50}MnO_{3+\delta}$, with the application of $10^4$ Oe magnetic fields, the suppression of thermal relaxation behavior is clearly evident.

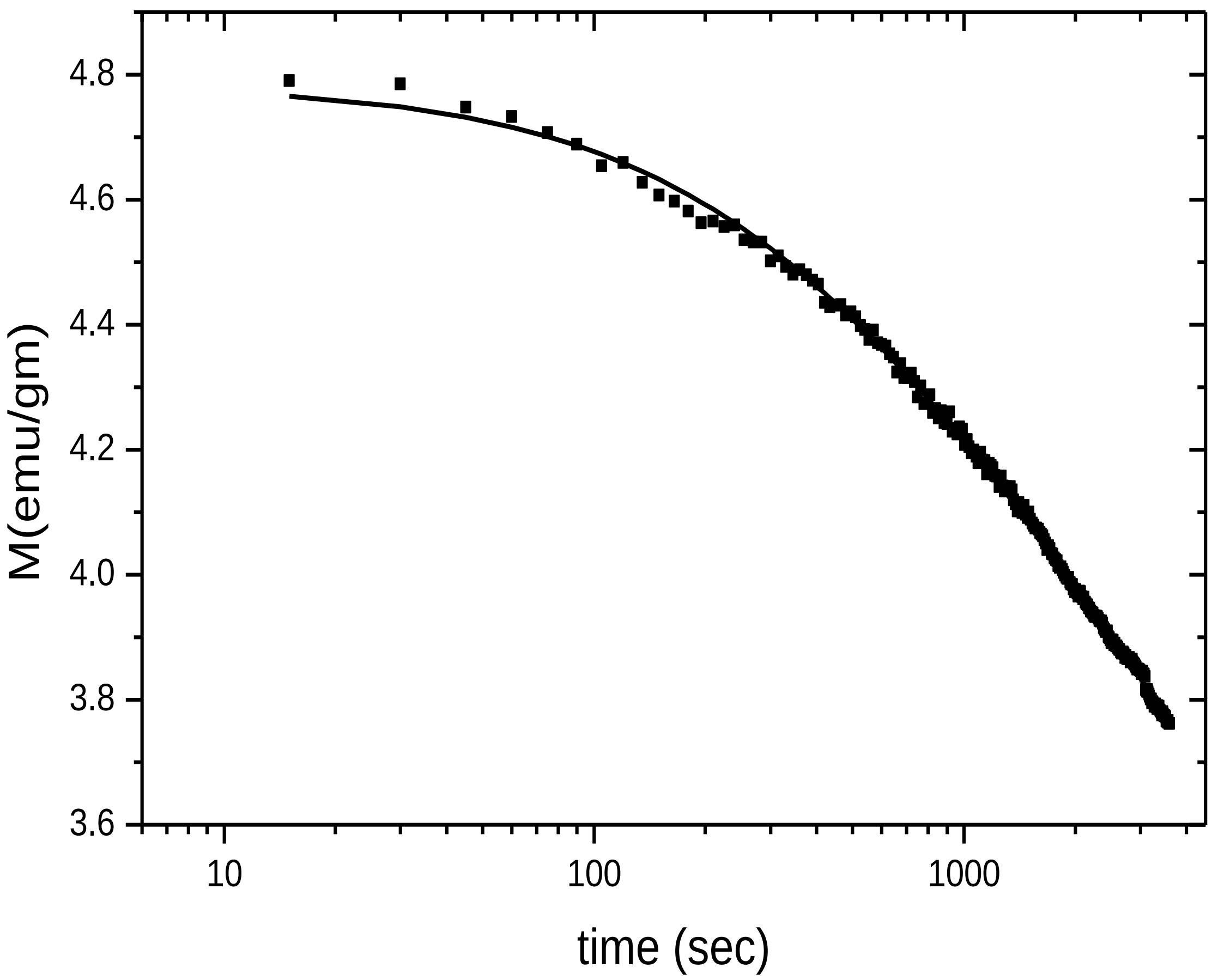

**Fig. 5.** DC magnetization as a function of time (DC relaxation) for the $La_{0.50}Ca_{0.50}MnO_{3+\delta}$ sample at T=80 K. The sample was field cooled under 50 Oe field but the sample was fresh and that has not undergone any thermal cycling. The solid line is fit to the $M=M_o(1-r\ln(1+(t/\tau))$ form as discussed in the text.

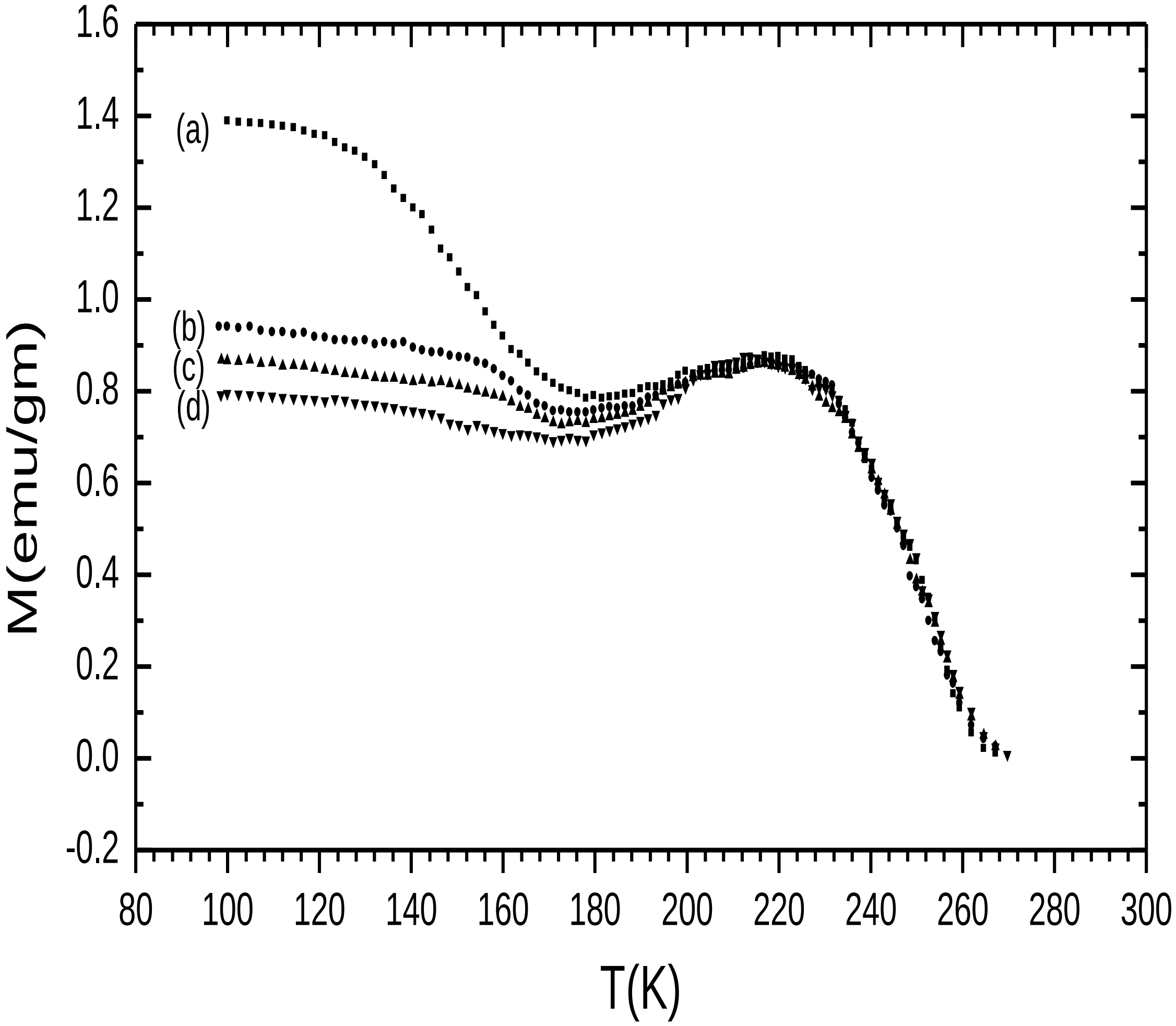

**Fig. 6.** The Zero Field Cooled (ZFC) DC magnetization (H=100 Oe) as a function of temperature for $La_{0.48}Ca_{0.52}MnO_{3+\delta}$ is shown. Curve (a) is for the as-prepared sample. Curve (b) was taken after two thermal cycles and then (c) and (d) were taken consecutively. Thermal cycling effects are clearly evident.

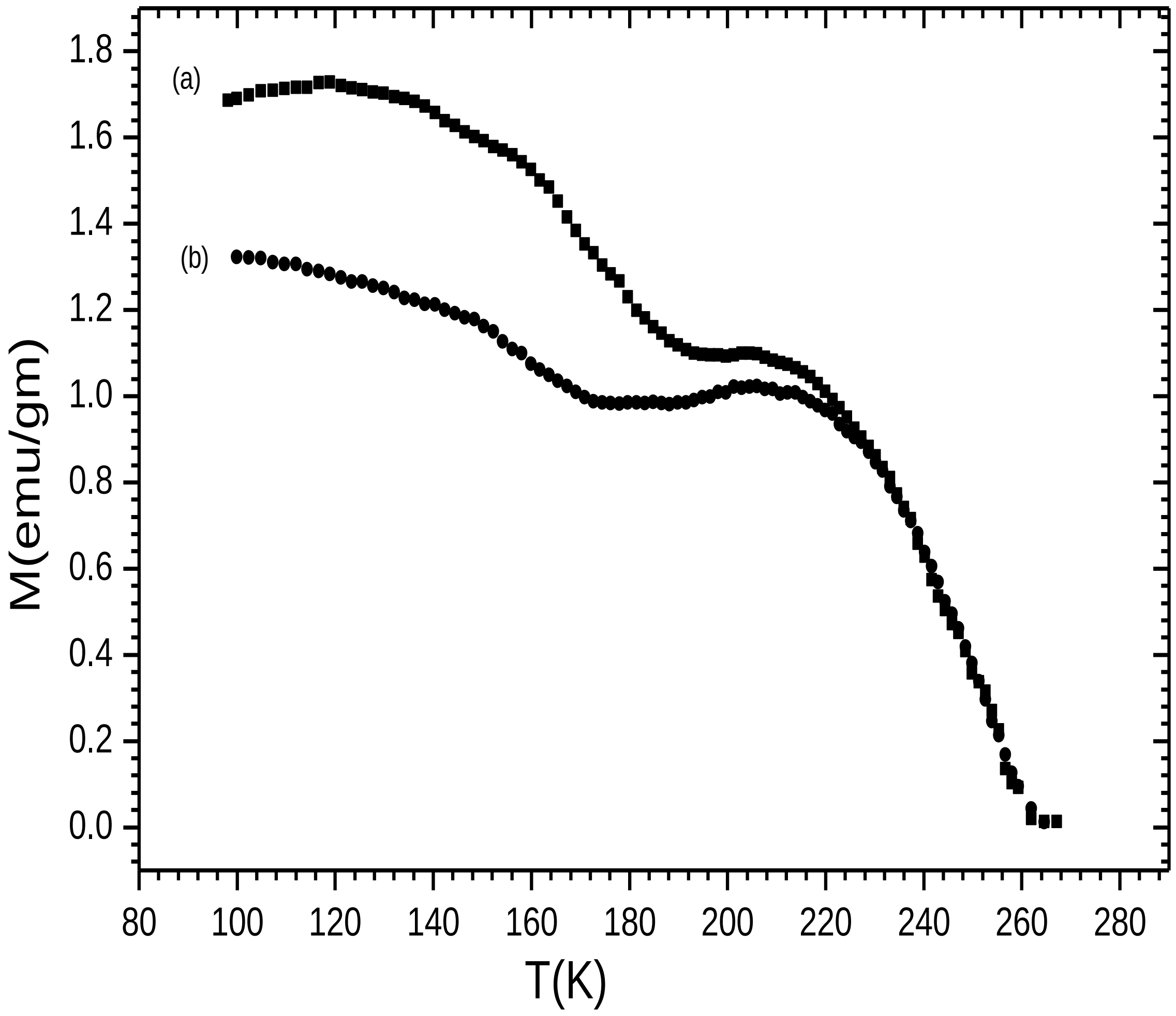

**Fig. 7.** The magnetization in the Field cooled (FC) mode (H=100 Oe) as a function of temperature for LCM-52 sample is shown. There were several thermal cycles in between the two curves (a) and (b).

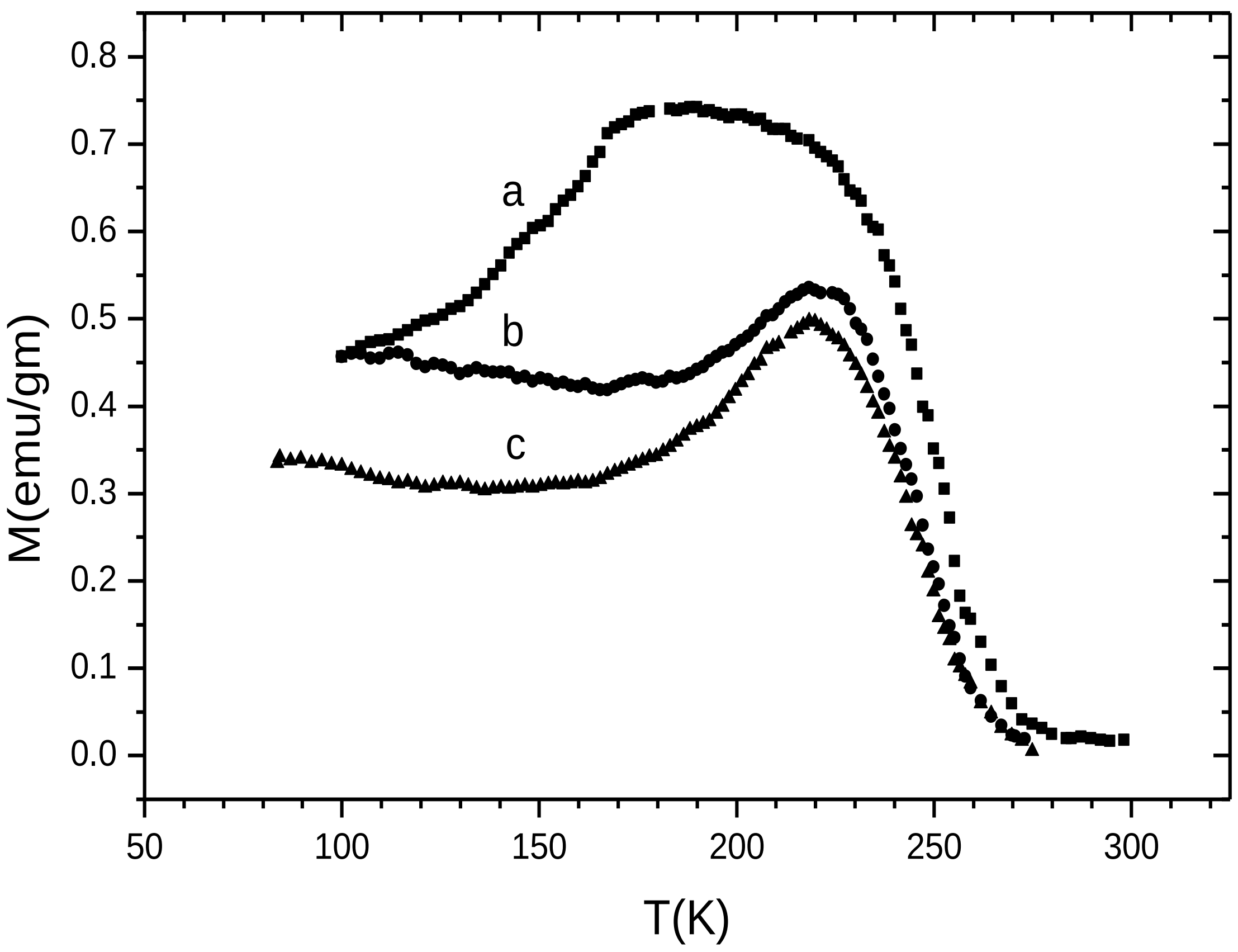

**Fig. 8.** Magnetization as a function of temperature is shown in the figure for LCM-55. Curves (a) and (b) were taken in cooling and heating the sample respectively in 100 Oe DC magnetic field. Curve (c) represents the ZFC magnetization for the same sample.

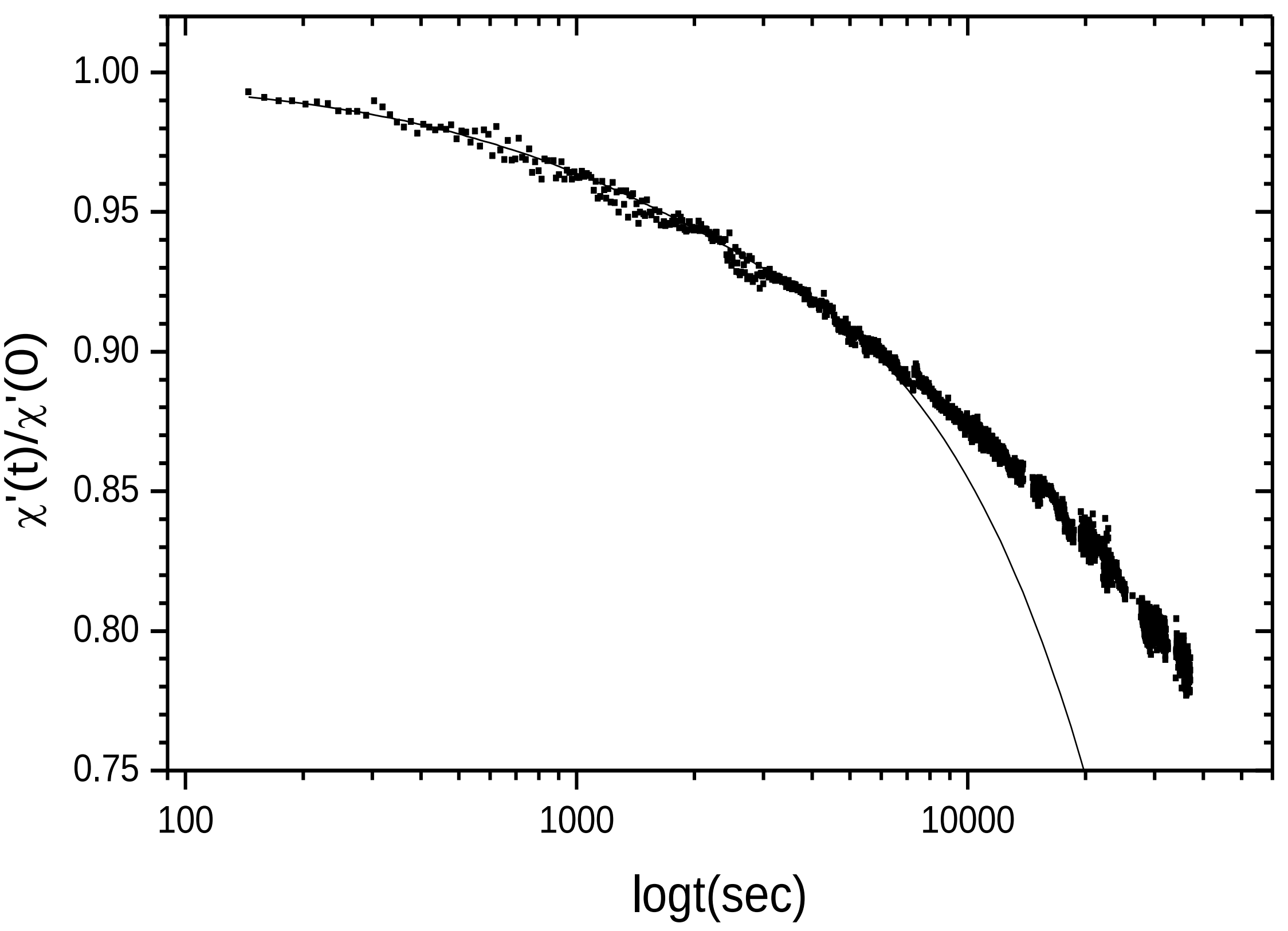

**Fig.9:** Typical time dependence of the AC susceptibility for the $La_{0.50}Ca_{0.50}MnO_{3+\delta}$ sample. The data are for the as prepared samples recorded for a period of one hour at T=80 K (left panel) and at T=90 K (right panel). The fit is to the ln(1+(T/τ)) form as discussed in the text. The rate of relaxation decreases with the increase in temperature.

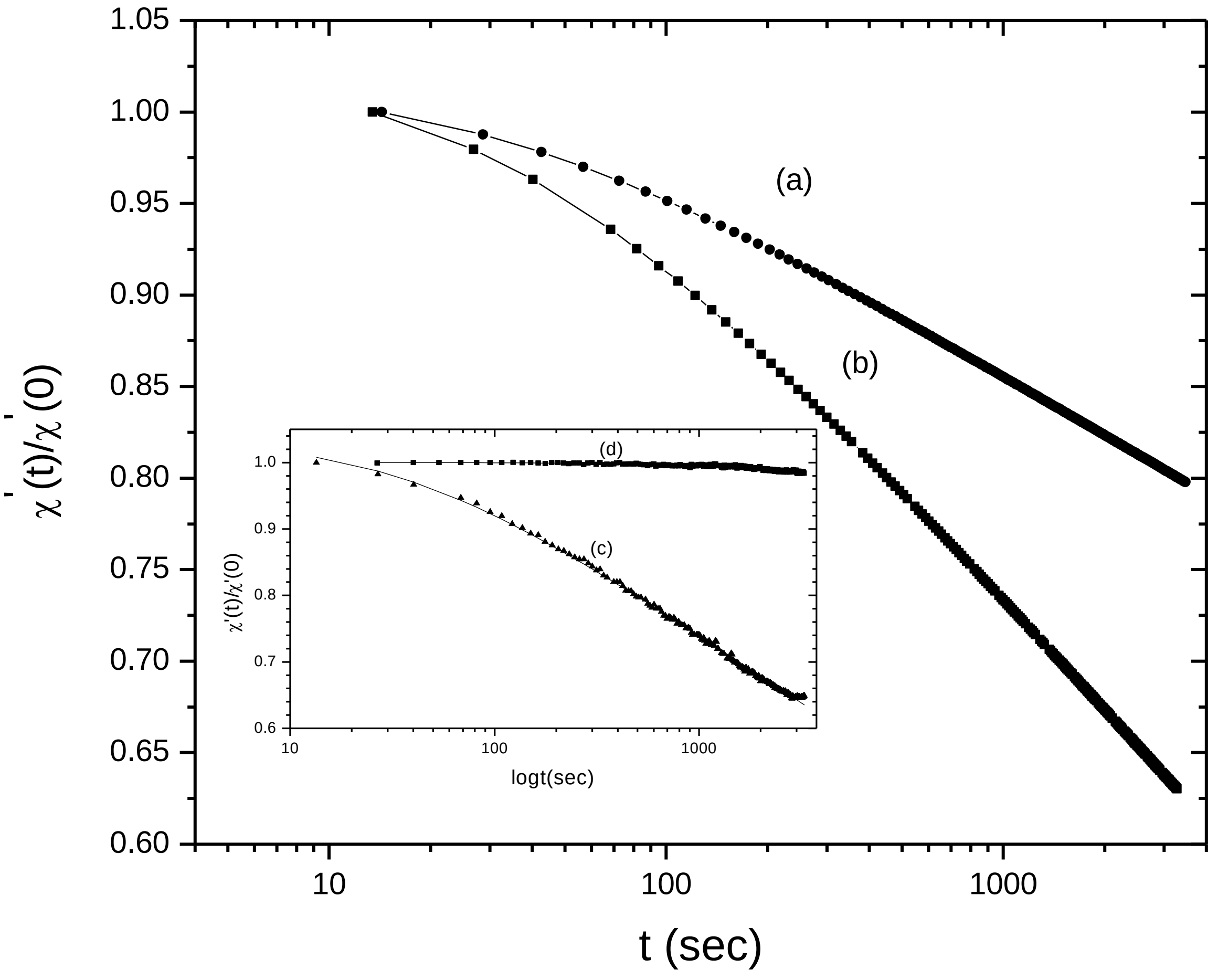

**Fig.10:** Temperature variation of the AC susceptibility of the $La_{0.50}Ca_{0.50}MnO_{3+\delta}$ sample for 5 Oe AC field and frequency f=131 Hz. Inset: Thermal cycling effects for the sample prior to annealing. The stabilization of the moment is evident between curves d and e. Main figure: Curve (a) was taken after annealing the same sample at T=700°C whereas curve (b) was taken subsequent to (a). The reactivation of the relaxation effects after annealing of the sample is demonstrated.